\input harvmac   
\noblackbox    

\def\p{\partial}

\font\cmss=cmss10    
\font\cmsss=cmss10 at 7pt    
\def\IL{\relax{\rm I\kern-.18em L}}    
\def\IH{\relax{\rm I\kern-.18em H}}    
\def\IR{\relax{\rm I\kern-.18em R}}    
\def\inbar{\vrule height1.5ex width.4pt depth0pt}    
\def\IC{\relax\hbox{$\inbar\kern-.3em{\rm C}$}}    
\def\rlx{\relax\leavevmode}    
\def\ZZ{\rlx\leavevmode\ifmmode\mathchoice{\hbox{\cmss Z\kern-.4em Z}}    
 {\hbox{\cmss Z\kern-.4em Z}}{\lower.9pt\hbox{\cmsss Z\kern-.36em Z}}    
 {\lower1.2pt\hbox{\cmsss Z\kern-.36em Z}}\else{\cmss Z\kern-.4em    
 Z}\fi}     
\def\IZ{\relax\ifmmode\mathchoice    
{\hbox{\cmss Z\kern-.4em Z}}{\hbox{\cmss Z\kern-.4em Z}}    
{\lower.9pt\hbox{\cmsss Z\kern-.4em Z}}    
{\lower1.2pt\hbox{\cmsss Z\kern-.4em Z}}\else{\cmss Z\kern-.4em    
Z}\fi}

\def\CC {{\cal C}}    
\def\CB {{\cal B}}    
    
\def\CA{{\cal A}}    
\def\CC{{\cal C}}   
    

\def\zb {\bar{z}}

\def\G{\Gamma}    
\font\manual=manfnt     
\def\dbend{\lower3.5pt\hbox{\manual\char127}}

\def\IZ{\relax\ifmmode\mathchoice    
{\hbox{\cmss Z\kern-.4em Z}}{\hbox{\cmss Z\kern-.4em Z}}    
{\lower.9pt\hbox{\cmsss Z\kern-.4em Z}} {\lower1.2pt\hbox{\cmsss    
Z\kern-.4em Z}}\else{\cmss Z\kern-.4em Z}\fi}    
\def\half {{1\over 2}}

\def\p{\partial}    
    
\def\bar{\overline}

\def\rt2{\sqrt{2}}    
\def\irt2{{1\over\sqrt{2}}}

\def\t{\tilde}    
    
\def\s{\sigma}    
\def\b{\beta}    
    
\Title{\vbox{\baselineskip12pt\hbox{hep-th/0406106}%
\hbox{HRI-P-0406001}%
\hbox{PUPT-2122}%
\hbox{TIFR/TH/04-15}%
\hbox{UT-04-17}}}%
{\vbox{\centerline{Liouville D-branes in Two-Dimensional Strings} 
\vskip8pt  
\centerline{and Open String Field Theory}}}  
  
{\vskip -20pt\baselineskip 14pt   
  
\centerline{Debashis Ghoshal$^{a,1,\dagger}$  
\footnote{}{$^\dagger$On sabbatical   
leave from the Harish-Chandra Research Institute, Chhatnag Road,   
Allahabad 211019, India.}%
\footnote{}{$^1$\tt ghoshal@hep-th.phys.s.u-tokyo.ac.jp},  
Sunil Mukhi$^{b,2}$ \footnote{}{$^2$\tt mukhi@tifr.res.in}  
and  
Sameer Murthy$^{c,3}$ \footnote{}{$^3$\tt smurthy@princeton.edu} }  
  
\bigskip  
\centerline{\sl $^a$Department of Physics, University of Tokyo,  
Tokyo, Japan}  
\smallskip  
\centerline{\sl $^b$Department of Theoretical Physics}  
\centerline{\sl Tata Institute of Fundamental Research,   
Mumbai, India}  
\smallskip  
\centerline{\sl $^c$Department of Physics, Princeton University,  
Princeton, NJ 08544, USA.}  
\bigskip\bigskip  
  
\noindent   
We study open strings in the noncritical $c=1$ bosonic string theory
compactified on a circle at self-dual radius. These strings live on
D-branes that are extended along the Liouville direction ({\it FZZT}
branes). We present explicit expressions for the disc two- and
three-point functions of boundary operators in this theory, as well as
the bulk-boundary two-point function. The expressions obtained are
divergent because of resonant behaviour at self-dual radius. However,
these can be regularised and renormalized in a precise way to get
finite results.  The boundary correlators are found to depend only on
the differences of boundary cosmological constants, suggesting a
fermionic behaviour. We initiate a study of the open-string field
theory localised to the physical states, which leads to an interesting
matrix model.
  
}  
\Date{June 2004}  
  
\lref\ShenkerNP{  
S.H.~Shenker,  
``The Strength Of Nonperturbative Effects In String Theory'',  
in the Proceedings of the Cargese Workshop on Random Surfaces,   
Quantum Gravity and Strings, (1990). }  
   
\lref\fzz{   
V.~Fateev, A.B.~Zamolodchikov and Al.B.~Zamolodchikov,    
``Boundary Liouville field theory. I: Boundary state and boundary  
two-point function,''   
arXiv:hep-th/0001012.   
}   
    
\lref\TeschnerThree{  
J.~Teschner,  
``On the Liouville three point function,''  
Phys.\ Lett.\ B {\bf 363}, 65 (1995)  
[arXiv:hep-th/9507109].  
}  
  
\lref\TeschnerRmk{  
J.~Teschner,  
``Remarks on Liouville theory with boundary,''  
arXiv:hep-th/0009138.  
}  
  
\lref\ZZPseud{  
A.B.~Zamolodchikov and Al.B.~Zamolodchikov,  
``Liouville field theory on a pseudosphere,''  
arXiv:hep-th/0101152.  
}  
  
\lref\TeschnerRV{  
J.~Teschner,  
``Liouville theory revisited,''  
Class.\ Quant.\ Grav.\  {\bf 18}, R153 (2001)  
[arXiv:hep-th/0104158].  
}  
   
\lref\hosomichi{   
K.~Hosomichi,   
``Bulk-boundary propagator in Liouville theory on a disc,''   
JHEP {\bf 0111}, 044 (2001)   
[arXiv:hep-th/0108093].   
}   
  
\lref\PonsTeschthree{  
B.~Ponsot and J.~Teschner,  
``Boundary Liouville field theory: Boundary three point function,''  
Nucl.\ Phys.\ B {\bf 622}, 309 (2002)  
[arXiv:hep-th/0110244].}  
  
\lref\kospon{    
I.~K.~Kostov, B.~Ponsot and D.~Serban,    
``Boundary Liouville theory and 2D quantum gravity,''    
[arXiv:hep-th/0307189].    
}    
  
\lref\teschner{    
J.~Teschner,    
``On boundary perturbations in Liouville theory and brane dynamics in    
noncritical string theories,''    
[arXiv:hep-th/0308140].    
}    
  
\lref\DornOtto{  
H.~Dorn and H.J.~Otto,  
``Two and three point functions in Liouville theory,''  
Nucl.\ Phys.\ B {\bf 429}, 375 (1994)  
[arXiv:hep-th/9403141].  
}  
  
\lref\ZZClosed{  
A.~B.~Zamolodchikov and A.~B.~Zamolodchikov,  
``Structure constants and conformal bootstrap in Liouville field theory,''  
Nucl.\ Phys.\ B {\bf 477}, 577 (1996)  
[arXiv:hep-th/9506136].  
}  
  
\lref\McV{  
J.~McGreevy and H.~Verlinde,  
``Strings from tachyons: The c = 1 matrix reloaded,''  
JHEP {\bf 0312}, 054 (2003)  
[arXiv:hep-th/0304224].  
}  
  
\lref\kms{   
I.~R.~Klebanov, J.~Maldacena and N.~Seiberg,   
``D-brane decay in two-dimensional string theory,''   
JHEP {\bf 0307}, 045 (2003)   
[arXiv:hep-th/0305159].   
}   
  
\lref\McTV{  
J.~McGreevy, J.~Teschner and H.~Verlinde,  
``Classical and quantum D-branes in 2D string theory,''  
JHEP {\bf 0401}, 039 (2004)  
[arXiv:hep-th/0305194].  
}  
  
\lref\newhat{  
M.R.~Douglas, I.R.~Klebanov, D.~Kutasov, J.~Maldacena, E.~Martinec   
and N.~Seiberg,  
``A new hat for the c = 1 matrix model,''  
arXiv:hep-th/0307195.  
}  
   
\lref\McMV{  
J.~McGreevy, S.~Murthy and H.~Verlinde,  
``Two-dimensional superstrings and the supersymmetric matrix model,''  
JHEP {\bf 0404}, 015 (2004)  
[arXiv:hep-th/0308105].  
}  
  
\lref\gaiotto{   
D.~Gaiotto and L.~Rastelli,   
``A paradigm of open/closed duality: Liouville D-branes and the Kontsevich   
model,''   
arXiv:hep-th/0312196.   
}   
  
\lref\NakaRev{  
Y.~Nakayama,  
``Liouville field theory: A decade after the revolution,''  
arXiv:hep-th/0402009.  
}  
  
\lref\SenOCone{  
A.~Sen,  
``Open-closed duality at tree level,''  
Phys.\ Rev.\ Lett.\  {\bf 91}, 181601 (2003)  
[arXiv:hep-th/0306137].  
}  
  
\lref\SenOCtwo{  
A.~Sen,  
``Open-closed duality: Lessons from matrix model,''  
Mod.\ Phys.\ Lett.\ A {\bf 19}, 841 (2004)  
[arXiv:hep-th/0308068].  
}  
  
\lref\SenOCthree{  
A.~Sen,  
``Rolling tachyon boundary state, conserved charges and   
two-dimensional string theory,''  
arXiv:hep-th/0402157.  
}  
 
\lref\KlebanovQA{ 
I.~R.~Klebanov, 
``String theory in two-dimensions,'' 
arXiv:hep-th/9108019. 
} 
 
\lref\DavidHJ{ 
F.~David, 
``Conformal Field Theories Coupled To 2-D Gravity In The Conformal Gauge,'' 
Mod.\ Phys.\ Lett.\ A {\bf 3}, 1651 (1988) \semi 
%
J.~Distler and H.~Kawai, 
``Conformal Field Theory And 2-D Quantum Gravity Or Who's Afraid Of Joseph 
Liouville?,'' 
Nucl.\ Phys.\ B {\bf 321}, 509 (1989). 
} 
 
\lref\DiFrancescoSS{ 
P.~Di Francesco and D.~Kutasov, 
``Correlation functions in 2-D string theory,'' 
Phys.\ Lett.\ B {\bf 261}, 385 (1991). 
} 
 
\lref\SakaiTF{
N.~Sakai and Y.~Tanii,
``Correlation functions of c=1 matter coupled to two-dimensional gravity,''
Prog.\ Theor.\ Phys.\  {\bf 86}, 547 (1991).
}

\lref\Penner{ 
R.~Penner, 
``Perturbative series and the moduli space of Riemann surfaces'',  
J.\ Diff.\ Geom. {\bf 27} (1988) 35. 
} 
 
\lref\DistlerMT{ 
J.~Distler and C.~Vafa, 
``A Critical Matrix Model At $c = 1$,'' 
Mod.\ Phys.\ Lett.\ A {\bf 6}, 259 (1991). 
} 
  
\lref\Distler{  
J.~Distler,  
``2-D Quantum Gravity, Topological Field Theory And The Multicritical Matrix  
Models,''  
Nucl.\ Phys.\ B {\bf 342}, 523 (1990).  
}  
  
\lref\Kontsev{  
M.~Kontsevich,  
``Intersection Theory On The Moduli Space Of Curves And The Matrix Airy  
Function,''  
Commun.\ Math.\ Phys.\  {\bf 147}, 1 (1992).  
}  
  
\lref\wittenCS{   
E.~Witten,   
``Chern-Simons gauge theory as a string theory,''   
Prog.\ Math.\  {\bf 133}, 637 (1995)   
[arXiv:hep-th/9207094].   
}   
  
\lref\openloc{  
A.~Kapustin and Y.~Li,  
``Topological correlators in Landau-Ginzburg models with boundaries,''  
Adv.\ Theor.\ Math.\ Phys.\  {\bf 7}, 727 (2004)  
[arXiv:hep-th/0305136];\hfill\break  
M.~Herbst, C.I.~Lazaroiu and W.~Lerche,  
``Superpotentials, A(infinity) relations and WDVV equations for open  
topological strings,''  
arXiv:hep-th/0402110;\hfill\break  
M.~Herbst and C.I.~Lazaroiu,  
``Localization and traces in open-closed topological Landau-Ginzburg 
models,''  
arXiv:hep-th/0404184.  
}  
   
\lref\sunilrev{   
S.~Mukhi,   
``Topological matrix models, Liouville matrix model and $c = 1$ string   
theory,''   
arXiv:hep-th/0310287.   
}   
 
\lref\HoravaCD{ 
P.~Horava, 
``Two-dimensional string theory and the topological torus,'' 
Nucl.\ Phys.\ B {\bf 386}, 383 (1992) 
[arXiv:hep-th/9202008]. 
} 
   
\lref\sunilvafa{   
S.~Mukhi and C.~Vafa,   
``Two-dimensional black hole as a topological coset model of $c = 1$   
string theory,''   
Nucl.\ Phys.\ B {\bf 407}, 667 (1993)   
[arXiv:hep-th/9301083].   
}   
   
\lref\tLG{  
D.~Ghoshal and S.~Mukhi,  
``Topological Landau-Ginzburg model of two-dimensional string theory,''  
Nucl.\ Phys.\ B {\bf 425}, 173 (1994)  
[arXiv:hep-th/9312189];\hfill\break  
A.~Hanany, Y.~Oz and M.~Ronen Plesser,  
``Topological Landau-Ginzburg formulation and integrable structure of 2-d  
string theory,''  
Nucl.\ Phys.\ B {\bf 425}, 150 (1994)  
[arXiv:hep-th/9401030];\hfill\break  
D.~Ghoshal, C.~Imbimbo and S.~Mukhi,  
``Topological 2-D string theory: Higher genus amplitudes and $W_\infty$  
identities,''  
Nucl.\ Phys.\ B {\bf 440}, 355 (1995)  
[arXiv:hep-th/9410034].  
}  
 
\lref\KostovCY{ 
I.~K.~Kostov, 
``Boundary ground ring in 2D string theory,'' 
Nucl.\ Phys.\ B {\bf 689}, 3 (2004) 
[arXiv:hep-th/0312301]. 
} 
  
\lref\DMP{  
R.~Dijkgraaf, G.W.~Moore and R.~Plesser,  
``The Partition function of 2-D string theory,''  
Nucl.\ Phys.\ B {\bf 394}, 356 (1993)  
[arXiv:hep-th/9208031].  
}  
  
\lref\IM{  
C.~Imbimbo and S.~Mukhi,  
``The Topological matrix model of $c = 1$ string,''  
Nucl.\ Phys.\ B {\bf 449}, 553 (1995)  
[arXiv:hep-th/9505127].  
}  
  
\lref\LZ{  
B.H.~Lian and G.J.~Zuckerman,  
``2-D Gravity With C = 1 Matter,''  
Phys.\ Lett.\ B {\bf 266}, 21 (1991).  
}  
  
\lref\MMS{  
S.~Mukherji, S.~Mukhi and A.~Sen,  
``Null vectors and extra states in $c = 1$ string theory,''  
Phys.\ Lett.\ B {\bf 266}, 337 (1991).  
}  
  
\lref\GRing{  
E.~Witten,  
``Ground ring of two-dimensional string theory,''  
Nucl.\ Phys.\ B {\bf 373}, 187 (1992)  
[arXiv:hep-th/9108004];\hfill\break  
E.~Witten and B.~Zwiebach,  
``Algebraic structures and differential geometry in 2-D string theory,''  
Nucl.\ Phys.\ B {\bf 377}, 55 (1992)  
[arXiv:hep-th/9201056].  
}  
  
\lref\debvafa{   
D.~Ghoshal and C.~Vafa,   
``$c = 1$ string as the topological theory of the conifold,''   
Nucl.\ Phys.\ B {\bf 453}, 121 (1995)   
[arXiv:hep-th/9506122].   
}   
   
\lref\wzwbrane{ 
V.~Schomerus, 
``Lectures on branes in curved backgrounds,'' 
Class.\ Quant.\ Grav.\  {\bf 19}, 5781 (2002) 
[arXiv:hep-th/0209241] \semi 
A.Y.~Alekseev and V.~Schomerus, 
``D-branes in the WZW model,'' 
Phys.\ Rev.\ D {\bf 60}, 061901 (1999) 
[arXiv:hep-th/9812193] \semi 
K.~Gawedzki, 
``Conformal field theory: A case study,'' 
[arXiv:hep-th/9904145] \semi 
S.~Stanciu, 
``D-branes in group manifolds,'' 
JHEP {\bf 0001}, 025 (2000) 
[arXiv:hep-th/9909163]. 
} 
 
\lref\Recknagel{ 
A.~Recknagel and V.~Schomerus, 
``Boundary deformation theory and moduli spaces of D-branes,'' 
Nucl.\ Phys.\ B {\bf 545}, 233 (1999) 
[arXiv:hep-th/9811237]. 
} 
 
\lref\TayZw{ 
W.~Taylor and B.~Zwiebach, 
``D-branes, tachyons, and string field theory,'' 
arXiv:hep-th/0311017. 
} 

\lref\AlexandrovNN{
S.~Y.~Alexandrov, V.~A.~Kazakov and D.~Kutasov,
``Non-perturbative effects in matrix models and D-branes,''
JHEP {\bf 0309}, 057 (2003)
[arXiv:hep-th/0306177].
}

\lref\MartinecKA{
E.~J.~Martinec,
``The annular report on non-critical string theory,''
arXiv:hep-th/0305148.
}

\lref\DeWolfeQF{
O.~DeWolfe, R.~Roiban, M.~Spradlin, A.~Volovich and J.~Walcher,
``On the S-matrix of type 0 string theory,''
JHEP {\bf 0311}, 012 (2003)
[arXiv:hep-th/0309148].
}

\lref\SeibergNM{
N.~Seiberg and D.~Shih,
``Branes, rings and matrix models in minimal (super)string theory,''
JHEP {\bf 0402}, 021 (2004)
[arXiv:hep-th/0312170].
}

\lref\TopoFermi{
M.~Aganagic, A.~Klemm, M.~Marino and C.~Vafa,
``The topological vertex,''
arXiv:hep-th/0305132\semi
M.~Aganagic, R.~Dijkgraaf, A.~Klemm, M.~Marino and C.~Vafa,
``Topological strings and integrable hierarchies,''
arXiv:hep-th/0312085.
}

\lref\GRprivate{
D.~Gaiotto and L.~Rastelli, private communications.
}
  
 
\newsec{Introduction}  
Non-critical string theories, describing strings propagating in 
two dimensions or less, were instrumental in shaping our understanding  
of the behaviour of string theory beyond the perturbative regime. The   
${\cal O}(1/g_s)$ nonperturbative effect, so characteristic of  
D-branes, first emerged from the study of these systems\ShenkerNP.  
Only recently, thanks to the advancement in our understanding   
of boundary Liouville   
dynamics\refs{\fzz\TeschnerThree\TeschnerRmk\ZZPseud\TeschnerRV 
\hosomichi\PonsTeschthree\kospon\AlexandrovNN\MartinecKA%
\DeWolfeQF\SeibergNM{--}\teschner} (following  
earlier work \refs{\DornOtto,\ZZClosed}), has a physical understanding of  
the nonperturbative effects begun to  
emerge\refs{\McV\kms\McTV\newhat\McMV{--}\gaiotto}. (See the review  
\NakaRev\ for an exhaustive list of references.)  
     
In another development, the dynamics of tachyon condensation led Sen  
to propose a new duality between open and closed  
strings\refs{\SenOCone\SenOCtwo{--}\SenOCthree}. Noncritical string  
theories are likely to be ideally suited for understanding this  
duality and indeed they have already played an important role in the  
shaping of these ideas. Recently in an interesting work\gaiotto,  
Gaiotto and Rastelli applied this philosophy to Liouville theory  
coupled to $c=-2$ matter.  This system has certain topological  
symmetries\Distler\ constraining its dynamics. Using these symmetries  
the authors obtain the Kontsevich topological matrix model\Kontsev\  
describing the {\it closed-string} theory starting from the  
{\it open-string field theory}.  
  
Among the non-critical string theories, the theory of a single scalar 
field coupled to worldsheet gravity has perhaps the richest 
structure. The matter theory has central charge $c=1$, while the 
Liouville field with its central charge $c_L=25$ provides an 
interpretation as a {\it critical} string theory with two-dimensional 
target space. Closed strings in this background have been studied in 
the past from quite a few different angles: matrix quantum mechanics 
(see Ref.\refs\KlebanovQA\ and references therein), worldsheet 
conformal field theory\refs{\DavidHJ\DiFrancescoSS{--}\SakaiTF}, 
topological field theory\refs{\HoravaCD\sunilvafa{--}\tLG} and 
topological matrix models\refs{\Penner\DistlerMT\DMP{--}\IM} related to 
the moduli space of Riemann surfaces (see \sunilrev\ for a recent 
review). 
 
The $c=1$ closed-string theory has a marginal deformation,
corresponding to changing the radius $R$ of compactification of the
scalar field. At a particular value of this radius, $R=1$ in our
conventions, the theory is self-dual under T-duality and an $SU(2)
\times SU(2)$ symmetry gets restored as a result of which momentum and
winding modes become degenerate with each other.
 
In this paper, we will consider the open-string version of this 
two-dimensional string theory -- more precisely, a scalar field 
compactified at the self-dual radius on a worldsheet with the topology 
of a disc or an upper half plane, coupled to the Liouville 
mode. Various types of branes are possible depending on the choice of 
boundary condition on the fields. We will choose to work with 
(generalized) Neumann boundary conditions on the Liouville field 
$\varphi$. On the matter field $X$ we impose Dirichlet boundary 
conditions, as a result of which the brane is localised in $X$ and 
there are no momentum modes in that direction, only winding 
modes. Because the radius is self-dual, one can equally well impose 
Neumann boundary conditions in $X$ and then there are momentum but no 
winding modes. The physics is identical in the two cases. 
 
The resulting branes are stable and are known as FZZT
branes\refs{\fzz\TeschnerThree{--}\TeschnerRmk}.  We will compute the
two- and three-point disc correlation functions of the fields living
on the FZZT branes, as well as the bulk-boundary two-point function of
such fields with `bulk' fields.  The Liouville contributions to these
correlators are non-trivial and general expressions are available in
the literature\refs{\fzz,\PonsTeschthree,\hosomichi,\teschner}.  Some
of them are only known in the form of contour integrals. From the
point of view of these theories, the $c_L=25$ Liouville field coupled
to $c=1$ matter is at the `boundary' of the theories studied in
Refs.\refs{\fzz\TeschnerThree%
\TeschnerRmk\ZZPseud\TeschnerRV\hosomichi\PonsTeschthree\kospon%
\teschner\DornOtto--\ZZClosed}. 
In the specific case of interest, we take a careful limit to obtain
the desired correlators. In particular we are able to evaluate the
relevant contour integrals in our case, leading to expressions that
are much simpler and more explicit than those previously given in the
literature for the more general $c\le 1$ case.
 
The resulting expressions satisfy the expected consistency conditions
and other recursion relations. When the Liouville theory is combined
with matter, one gets a massless `tachyon' field\foot{No operator in
this paper is truly unstable on the worldsheet. The FZZT boundary
conditions do not allow such modes. With this in mind, and since there
is also no unstable operator in the bulk, we use the word tachyon as
is conventional.} labelled by integer winding numbers. The matter
contribution to the tachyon correlators are just
winding number conserving delta functions. In addition to the
tachyons, there are discrete states at ghost numbers one and
zero\refs{\LZ,\MMS}. The former are the remnants of massless and
massive states of critical strings and their correlators are
determined by the SU(2) symmetry at the self-dual radius. (The latter
class of operators are characteristic of non-critical theories and in
particular, they form a ring on which the symmetry of the theory can
be realized in a geometric way\GRing.) We have not attempted to study
this ring in the FZZT brane background (for general results on the
$c\le 1$ boundary ground ring, see Ref.\refs\KostovCY). As mentioned
above, the expressions for these correlators are divergent. As we will
see, once we perform renormalizations of the bulk and boundary
cosmological constants, the divergence is a common multiplicative
factor for both the two- and three-point boundary correlators.
 
The simple and elegant form of the answers obtained is suggestive of a
simple physical interpretation, perhaps in terms of fermions, as we
will see. The answers share some of the properties of the (simpler)
case of $c=-2$\refs\gaiotto, notably that they are independent of the
bulk cosmological constant. All this encourages us to try and
understand the corresponding open-string field theory, following the
ideas in Ref.\refs{\gaiotto}. Accordingly, in the last section of this
paper, we begin to study the open-string field theory of the FZZT
branes. Motivated by the fact that the disc path integral describing
classical processes of non-critical string theory localizes to the BRS
cohomology, we evaluate the action for the `on-shell' states (tachyons
and the discrete states).  This results in a non-local theory of
infinitely many matrices.  We hope to analyse this theory in more
detail in the future.
 
The organisation of this paper is as follows. Sec.~2 describes the
background and sets up notation. The Liouville contributions to the
two- and three-point functions are evaluated in Sections 3 and 4. In
Sec.~5 we calculate the bulk-boundary two-point function. Sec.~6 is
devoted to string field theory. We end with some comments in the final
Sec.~7.  Appendix A contains some properties of the special functions
that appear in the Liouville correlators. Some details of the contour
integral relevant for Sec.~4 are given in Appendix B.
 

\newsec{Two-dimensional Open String Theory and the FZZT branes}  
The theory we are interested in is described by the worldsheet  
action\foot{We work in $\alpha'=1$ units.} 
\eqn\wsaction{\eqalign{  
&{1\over 4\pi}\int_{\cal D}\left((\p X)^2 + (\p\varphi)^2 +   
Q\hat R\varphi + 4\pi\mu_0\, e^{2 b \varphi}\right)\cr  
+ &{1\over 2\pi}\int_{\p{\cal D}}\left(Q\hat K\varphi +   
2\pi\mu_{B,0}\, e^{b\varphi}\right),  
}}  
where $X,\phi$ are the matter and Liouville fields and $Q,b$ are 
numerical coefficients. In this action, ${\cal D}$ has the topology of 
a disc/UHP, $\hat R$ is the curvature of the (reference) metric, $\hat 
K$ the induced curvature of the boundary and $\mu_0$ and $\mu_{B,0}$ 
are the (bare) bulk and boundary cosmological constants respectively. 
 
With the action above, the matter sector has central charge $c=1$ 
while the Liouville sector has $c_L=1+6Q^2$. The coefficient $b$ 
appearing in the exponents satisfies $Q= b+{1\over b}$. 
Criticality requires the choice $c_L=25$, from which we determine 
$Q=2$ and $b =1$. Because of divergences that appear at $b=1$, we 
will need to carefully take the limit $b\to 1$ and regularise 
the divergences appropriately. 
 
On the field $\varphi$, we will impose  
\eqn\fzztbc{  
i(\p - \bar\p)\varphi = 4\pi\mu_{B,0}\, e^{b \varphi}, }  
the generalized Neumann boundary condition.  
  
The field $X$, which we take to be Euclidean is, in general,
compactified on a circle of radius $R$. We can impose a suitable
boundary condition on the field $X$; for instance, at a generic
radius, we could impose Dirichlet or Neumann boundary conditions $(\p
\pm \bar\p)X=0$ which, in conjunction with the boundary condition on
the Liouville field above, describe the non-compact $D$-instanton and
$D0$-brane respectively. As noted in the introduction, the two choices
are physically equivalent at self-dual radius.
 
In the bulk, the observables of the theory are a massless scalar field 
in the two dimensional target space known as the `tachyon' field, and 
an infinite set of quantum mechanical states which arise at special 
values of the momentum, known as the discrete states.  The vertex 
operators\foot{We are only considering local operators, which 
correspond to non-normalizable modes.} corresponding to the tachyon 
field take the following form at weak coupling: 
\eqn\bulktachvert{  
{\cal T}_k = c\bar c\,  
\exp\left(ik (X \pm \bar X) + (2-|k|)\varphi \right).  
}  

The tachyon vertex operator on the boundary on the other hand carries  
additional indices $(\s_1,\s_2)$ corresponding to the boundary  
conditions on the two ends of the open string:  
\eqn\bndrytach{  
T_k^{\s_1\s_2} \equiv c\,\left[e^{ikX}\, V_{\beta}\right]^{\s_1 \s_2}  
= c\,\left[\exp\left(ikX + \beta\varphi\right)\right]^{\s_1\s_2}, } 
where the second expression is the asymptotic form. {}From this we
see that $\beta$ labels the Liouville momentum, and the conformal
dimension of the Liouville vertex operator is $\Delta =
\beta(Q-\beta)$ where $Q=b + {1\over b}=2$. Requiring that the full
vertex operator has dimension one, one finds the on-shell condition
$\beta=1-|k|.$
The boundary label $\s$ is related to the  
(bare) cosmological constants $\mu_0$ and $\mu_{B,0}$ by:  
\eqn\boundarycc{    
\cos 2\pi b\left(\sigma - {Q\over 2}\right) =     
{\mu_{B,0}\over\sqrt{\mu_0}}\sqrt{\sin\pi b^2}. }    
As we shall discuss later, the cosmological constants require  
renormalisation in the $c=1$ string theory.  
 
We are specifically interested in the theory at self-dual radius
$R=1$, where the worldsheet theory is an $SU(2)_L \times SU(2)_R$
current algebra at level $1$.  The symmetry of the closed-string
theory is generated by $(J^{\pm}, J^3) = (e^{\pm i 2 X}, i \p X)$ and
their right moving counterparts.  The physical vertex operators at
ghost number one are\GRing:
\eqn\bulksutwovert{  
Z_{k;m,\bar m} = c\bar c\,  
V^{mat}(k,m){\bar V^{mat}(k,\bar m)}   
\exp\left((2-k)\varphi \right).  
} 
where $k$ is a non-negative integer or half integer; $V^{mat}(k,k)
\equiv e^{i k X }$, and the operators $V^{mat}(k,m<k)$ are defined by
acting with the $SU(2)_L$ lowering operator. Hence
$m=k,k-1,\cdots,-k$. The corresponding right movers are defined in a
similar manner. The physical content of the theory can also be
summarized as a massless field ${\cal T}(\theta,
\phi,\psi;\varphi)$ living on an $S^3$ times the  non-compact Liouville 
direction. 
 
The open string imposes a boundary condition relating the left and 
right moving currents $J^a$ and $\bar J^a$. The branes in the 
$SU(2)_n$ theory are labelled by a half-integer $J=0,\cdots,{n\over 
2}$ which labels the conjugacy class in the group, and continuous 
moduli which take values in $SO(3)$ which label the origin of the 
3-sphere viewed as a group manifold\wzwbrane . The conjugacy classes are  
topologically 2-spheres in the group manifold. 
 
For our case, level $n=1$, there are only two possible discrete labels 
$J=0,1/2$ and the full moduli space is $SU(2)$ which is topologically  
$S^3$\Recknagel. A brane is simply a point on this sphere, which can be 
thought of as a degenerate $S^2$. It breaks the $SO(4) = SU(2) \times 
SU(2)$ symmetry of the 3-sphere to a diagonal $SU(2)$ symmetry group 
of the degenerate 2-sphere.  The open-string modes are classified as 
representations of this $SU(2)$. 
  
For example, the boundary states which correspond to Neumann and 
Dirichlet for generic radii are labelled by the two poles on the  
$S^3$, and are given respectively by $J^a =  
\pm \bar J^a$. The generators of the diagonal $SU(2)$ subgroup which is  
preserved are $J^a\pm\bar J^a$. The allowed representations of the  
diagonal SU(2) are $k \pm\bar k$ where both $k$ and $\bar k$ are integer  
or half integer, so that the allowed representations of the diagonal  
subgroup are integer. Note that half the representations of $SU(2)$  
(the half-integer spins) do not correspond to physical operators.  
 
Thus the physical vertex operators of the open string at ghost number
one are:
\eqn\bdrysutwovert{  
Y^{\s_1\s_2}(k,m) \equiv c\, \left[V^{mat}(k,m)  
V_\beta\right]^{\s_1,\s_2} = c\, \left[V^{mat}(k,m) \exp((1-k)  
\varphi)\right]^{\s_1\s_2},} 
where, $(k,m)$ are the usual $SU(2)$ labels with spin $k$ an integer 
and $m=k,k-1,\ldots,-k$.  

 
\newsec{Boundary Two-Point Function}    
 
In this section, we shall compute the two-point function of the
Liouville vertex operators $V_\beta^{\s_1 \s_2}$ which enter the
physical open-string vertex operators \bdrysutwovert. The two-point
function of boundary operators in Liouville theory, of arbitrary
central charge $c_L=1+6Q^2$, is given by\fzz:
\eqn\twoptbasic{    
\left\langle V_{\b_1}^{\s_1 \s_2}(x)V_{\b_2}^{\s_2 \s_1}(0)\right\rangle   
\equiv     
{\delta(\b_1+\b_2-Q) + d(\beta|\sigma_1, \sigma_2)\delta(\b_1-\b_2)   
\over |x|^{2\Delta_{\b_1}}},  }    
where $d(\beta|\sigma_1, \sigma_2)$ is the {\it reflection amplitude}, 
the expression for which is given below.  The delta functions can be 
understood as arising due to the reflection from the Liouville 
potential, and is not present in the higher-point functions. Every 
non-normalizable operator in the theory is related to a normalizable 
operator by this reflection, $V_\b^{\s_1 \s_2} = 
d(\beta|\sigma_1,\sigma_2) V_{Q-\b}^{\s_1 \s_2}$. 
 
The reflection amplitude $d(\beta|\sigma_1, \sigma_2)$ is given by\fzz: 
\eqn\twoptfactors{\eqalign{   
d(\beta|\sigma_1, \sigma_2) &= \CA_1\, \CA_2\, \CA_3, \cr   
\CA_1 &= \left(\pi \mu_0 \gamma(b^2)b^{2-2b^2}\right)^{{Q-2\beta\over   
2b}},\cr    
\CA_2 &= {\Gamma_b(2\beta-Q)\over \Gamma_b(Q-2\beta)},\cr    
\CA_3 &= {S_b(2Q-\sigma_1-\sigma_2-\beta)\,    
S_b(\sigma_1+\sigma_2 -\beta)    
\over S_b(\beta + \sigma_1 - \sigma_2) \,    
S_b(\beta - \sigma_1 + \sigma_2)}. }}     
In the above, $\gamma(x) \equiv \Gamma(x)/\Gamma(1-x)$ and the special 
functions $\Gamma_b(x)$ and $S_b(x)$ are defined in 
\refs{\fzz,\teschner}. We record the relevant details in Appendix A. 
    
As mentioned above, to specialise to $c=1$ we must carefully take the 
limit $b\to 1,Q\to 2$. This limit is singular and requires us first of 
all to renormalize both the bulk and boundary cosmological constants. 
In the first line in Eq.\twoptfactors, we set $b= 1-{\varepsilon 
\over 2}$ and find that    
\eqn\aoneren{    
\CA_1 \to \left(\pi\mu_0  \gamma(1-\varepsilon)\right)^{1-\beta}. }    
Using $\gamma(1-\varepsilon) \to \varepsilon$, we see that the above    
expression becomes finite if we define the renormalised\foot{This    
differs by a factor of 4 from the normalisation used in    
Refs.\refs{\kospon,\kms}. However, it is more natural as the area of a    
unit 2-sphere is $4\pi$.} bulk cosmological constant by:    
\eqn\muren{    
\mu = 4\pi \mu_0\, \epsilon.}    
Using this and recalling from \bdrysutwovert\ that $\beta=1-k$ with $k$  
a non-negative integer, it follows that the first factor in the two-point  
amplitude is:  
\eqn\aone{    
\CA_1= \left({\mu\over4}\right)^{1-\beta} = \left({\sqrt\mu\over    
2}\right)^{2k}. }    
The renormalisation of the bare bulk cosmological constant $\mu_0$    
performed above is well-known, and leads to the result that the    
cosmological operator for $c=1$ closed strings is not the naive one,     
$e^{2\varphi}$, but rather $\varphi\,e^{2\varphi}$.    
    
Now coming back to Eq.\boundarycc\ and taking $b=1-{\varepsilon\over 2}$    
we have, for small $\varepsilon$:    
\eqn\bccagain{    
\cos 2\pi \sigma = \sqrt{\pi\varepsilon}\, {\mu_{B,0}\over \sqrt{\mu_0}}    
= 2\pi\varepsilon\, {\mu_{B,0}\over \sqrt{\mu}}, }    
which means that we also need to define a renormalised\foot{Once again
this differs (now by a factor of 2) from the normalisations of
Refs.\refs{\kms,\kospon}, and is consistent with the length of a unit
circle being $2\pi$.} boundary cosmological constant $\mu_B =
2\pi\varepsilon\,\mu_{B,0}$.  Hence finally the relation between the
$\sigma$ parameter and the renormalised (bulk and boundary)
cosmological constants is:
\eqn\bccren{    
\cos 2\pi \sigma = {\mu_B\over \sqrt\mu}. }    
The parameter $\sigma$ can be real or imaginary depending on whether    
$\mu_B<\sqrt\mu$ or $\mu_B>\sqrt\mu$. In what follows, we keep all the   
$\sigma_i$ generic.    
    
The factor $\CA_2$ depends only on $\beta$ and not on    
$\sigma_i$. Using Eq.\twoptfactors, we find:    
\eqn\atwo{    
\CA_2 = {\Gamma_1(-2k)\over \Gamma_1(2k)}. }    
This expression is actually divergent. However, we can regulate it by
going slightly off-shell. We can do this by shifting $\beta$ from the
integer value by an amount $\epsilon$: $k \to k+\epsilon$ and extract
the leading divergence.  We could use a different regulator and deform
$b$ away from $1$ to $1-\epsilon$ and we get the same answer. As
detailed in Appendix A, ${\cal A}_2$ is determined to be:
\eqn\atwofinal{    
\CA_2 =  {(-1)^{k}\over (2\pi)^{2k}\,    
\Gamma(2k+1)\Gamma(2k)} {1\over\epsilon^{2k+1}}. }    

Finally we turn to the third factor in Eq.\twoptbasic:    
\eqn\facthree{    
\CA_3 =     
{S_1(2Q-\sigma_1-\sigma_2-\beta)\,    
S_1(\sigma_1+\sigma_2 -\beta)    
\over S_1(\beta + \sigma_1 - \sigma_2) \,    
S_1(\beta - \sigma_1 + \sigma_2)}. }  
Now using the inversion relation $S_b(x)S_b(Q-x)=1$ and substituting    
$\beta = 1-k$, $\CA_3$ can be rewritten as:    
\eqn\facthreenew{\eqalign{    
\CA_3 & = {S_1(\s_1+\s_2-\b) \over S_1(-Q+\b+\s_1+\s_2)}     
{S_1(Q-\b +\s_1-\s_2) \over S_1(\b + \s_1-\s_2)} \cr    
& =  {S_1(-1+k+\s_1+\s_2) \over S_1(-1-k+\s_1+\s_2)}     
{S_1(1+k+\s_1-\s_2) \over S_1(1-k + \s_1-\s_2)}. }}    
Next we define the combinations $\sigma^\pm = \sigma_1 \pm \sigma_2$   
and invoke the recursion relation (see Appendix A)   
$S_1(x+1) = 2\sin\pi x\,S_1(x)$    
to write:    
\eqn\recurresult{\eqalign{    
\CA_3 &= \prod_{m=1}^{2k} \left( 2\sin\pi(\sigma^+ + k -1-m)\right)    
\prod_{n=1}^{2k} \left( 2\sin\pi(\sigma^- + k -1-n)\right)\cr   
&= \left(4\,\sin\,\pi\sigma^+\;\sin\,\pi\sigma^-\right)^{2k}. }}    
This can be rewritten in terms of the original boundary parameters  
$\s_1$ and $\s_2$:  
\eqn\cathreefinal{    
\CA_3     
=\left(2\, (\cos{2\pi\s_1}-\cos{2\pi\s_2})\right)^{2k}    
=\left(2\,{\mu_{1B}-\mu_{2B} \over \sqrt{\mu}}\right)^{2k}. }     

Putting everything together, we finally get:    
\eqn\twopointfinal{    
d(1-k|\mu_{1B},\mu_{2B}) = {(-1)^{k} \over\epsilon^{2k+1}}    
{\left(\mu_{1B}-\mu_{2B}\right)^{2k}\over    
(2\pi)^{2k}\,\Gamma(2k+1)\Gamma(2k)}. }    
We will find it convenient to renormalize the open-string  
operators \bdrysutwovert\ as  
\eqn\renfield{    
\t Y^{\s_1,\s_2}_{k,m} = (2\pi \epsilon)^{k} \Gamma(2k)\,  
Y^{\s_1\s_2}_{k,m}.} 
This redefinition is different from the standard one found in the
literature for closed strings, as it has an additional factor of 
$(2\pi\epsilon)^k$.  For
the cosmological operator ($k=0$) this extra factor is absent and the
renormalisation is the standard one.  The matter contribution to the
two-point function being trivial, let us put the renormalization
factor in the Liouville vertex operator alone and define
\eqn\renLiou{ 
\t V_{1-k}^{\s_1,\s_2} = (2\pi \epsilon)^{k} \Gamma(2k)\,  
V_{1-k}^{\s_1,\s_2}. } 

Expressed in these variables, the reflection amplitude   
is:    
\eqn\rentwopt{    
{\tilde d}(1-k|\mu_{1B},\mu_{2B}) = {(-1)^{k} \over \epsilon}      
{\left(\mu_{2B}-\mu_{1B}\right)^{2k} \over 2k}. }    
Several features of this result are noteworthy.  First, it is
independent of the bulk cosmological constant $\mu$. A similar feature
was noticed\refs{\gaiotto} for the correlators of $c=28$ Liouville
theory (corresponding to strings propagating in a $c=-2$ matter
background).  Second, the result depends only on the difference of the
two boundary cosmological constants $\mu_{1B},\mu_{2B}$. We will see
later that these features persist for the boundary three-point
function. They are reminiscent of the identification of the extended
B-type branes of topological field theories to fermions\TopoFermi.
Finally, we see that after renormalization, the reflection amplitude
has a simple pole singularity (as a function of $\epsilon$).  Again
this turns out to be the case for the boundary three-point function as
well. Later, when we use this in the string field theory action, we
will need to absorb this singularity by a redefinition of the string
coupling constant.
    

\newsec{Boundary Three-Point Function} 
    
The three-point function in boundary Liouville theory is defined by:   
\eqn\threept{    
\langle  V^{\s_2\s_3}_{\beta_1}(x^1)   
V^{\s_3\s_1}_{\beta_2}(x^2)   
V^{\s_1\s_2}_{\beta_3}(x^3)\rangle   
=   
{C^{\s_2\s_3\s_1}_{\b_1\b_2\b_3}\over    
|x_{21}|^{\Delta_1 + \Delta_2-\Delta_3}   
|x_{32}|^{\Delta_2 + \Delta_3-\Delta_1}   
|x_{13}|^{\Delta_3 + \Delta_1-\Delta_2}}.  
}    
An expression was found in Ref.\refs{\PonsTeschthree} (see also 
Refs.\refs{\kospon,\teschner} for subsequent discussions) for the 
coefficient $C$ as a product of four factors: 
\eqn\defBall{\eqalign{    
C^{\s_2\s_3\s_1}_{\b_1\b_2\b_3}&= \CB_1\,\CB_2\,\CB_3\,\CB_4,  , \cr 
\CB_1  &= \left(\pi \mu \gamma(b^2)     
b^{2-2b^2}\right)^{\half (Q-\beta_1-\beta_2-\beta_3)},\cr    
\CB_2 &= {\G_b(\beta_2+\beta_3-\beta_1)   
\G_b(2Q-\beta_1-\beta_2-\beta_3)     
\G_b(Q-\beta_1-\beta_2+\beta_3)   
\G_b(Q-\beta_1+\beta_2-\beta_3)    
\over    
\G_b(Q)\, \G_b(Q-2\beta_1)\, \G_b(Q-2\beta_2)\, \G_b(Q-2\beta_3) },\cr    
\CB_3 &=     
{S_b(Q-\beta_3+\sigma_1-\sigma_3)\, S_b(2Q-\beta_3-\sigma_1-\sigma_3) 
\over  S_b(\beta_2+\sigma_2-\sigma_3)\,    
S_b(Q+\beta_2-\sigma_2-\sigma_3)},\cr    
\CB_4 &= {1 \over i} \int_{-i \infty -0}^{+i \infty -0} ds \;\;    
\prod_{i=1}^4{S_b(U_i+s) \over S_b(V_i+s)}.    
} }    
In the factor $\CB_4$, the quantities $U_i, V_i$, $i=1,\cdots,4$ are    
defined as follows:    
\eqn\defuv{\eqalign{    
U_1 =\sigma_1+\sigma_2-\beta_1, \qquad  & V_1 = 2Q+    
\sigma_2-\sigma_3-\beta_1-\beta_3, \cr    
U_2 = Q-\sigma_1+\sigma_{2}-\beta_1, \qquad & V_2 = Q+\sigma_2-    
\sigma_3-\beta_1+\beta_3, \cr     
U_3 = \beta_2+\sigma_2-\sigma_3,  \qquad & V_3 = 2\sigma_2, \cr    
U_4 = Q-\beta_2+\sigma_2-\sigma_3, \qquad & V_4=Q.   
}}    
We want to compute the above for the values $b=1$, $\beta_i = 1-k_i$
for our case of $c=1$. In this section, we choose the kinematic regime
$k_3 > k_1,k_2 > 0$.  We shall later need to take a careful limit as
$k_i$ approach integers. The first two factors are evaluated as
before, and we get
\eqn\Bonetwo{   
\eqalign{    
\CB_1 &=  \left({\mu\over 4}\right)^{\half(k_1+k_2+k_3-1)}    
= \left({\sqrt\mu\over 2}\right)^{\sum_i k_i -1},\cr    
\CB_2 &= { \G_1(1+\sum_i k_i)\G_1(1+k_1+k_2-k_3)  
\G_1(1+k_1-k_2+k_3)\G_1(1+k_1-k_2-k_3)   
\over     
\G_1(2)\, \G_1(2k_1)\,  \G_1(2k_3)\, \G_1(2k_2)}\cr    
& =  {(-1)^{\lfloor(k_2+k_3-k_1)/2\rfloor} \over   
(2\pi \epsilon)^{k_2+k_3-k_1}}  
\; \times {\G_1\left(1+\sum_i k_i \right)\over \G_1(2)} \cr  
& \qquad\times\;   
{\G_1(1+k_1+k_2-k_3)\, \G_1(1+k_1-k_2+k_3)\, \G_1(1-k_1+k_2+k_3)  
\over  \G_1(2k_1)\,  \G_1(2k_2)\, \G_1(2k_3)} \cr  
& = {(-1)^{\lfloor(k_2+k_3-k_1)/2\rfloor} \over   
(2\pi \epsilon)^{k_2+k_3-k_1}}  
{\G_1\left(1+\sum_i k_i \right) \over \G_1(2)}   
\prod_{j=1}^3  
{\G_1\left(1+\sum_i k_i - 2k_j \right) \over \G_1(2k_j)}, 
}}    
where $\lfloor x\rfloor$ is the integer part of $x$.  We have used the
properties of the special function $\G_1(x)$ at integer arguments
given in Appendix A to rewrite the last factor in the numerator of
$\CB_2$.
 
For the factor $\CB_3$, we insert the values of the parameters to    
write it as:     
\eqn\Bthree{\eqalign{    
\CB_3 = {S_1(1+k_1+k_2+\sigma_1-\sigma_3)\,     
S_1(3+k_1+k_2-\sigma_1-\sigma_3)     
\over S_1(1-k_2+\sigma_2-\sigma_3)\, S_1(3-k_2-\sigma_2-\sigma_3)}.    
}}    
It turns out that this simplifies when combined with a similar    
factor in $\CB_4$.    
    
Finally we must evaluate the contribution $\CB_4$. This is carried out
in Appendix B, where the contour integral in the last line of
Eq.\defBall\ is evaluated explicitly. That is then combined with
$\CB_3$ of Eq.\Bthree\ above to give the following amazingly simple
result for the product:
\eqn\BthreeBfour{    
\CB_3 \,\CB_4 =   {(-1)^{k_1} \over (2 \pi \epsilon)^{2k_1+1}}     
\left({2\mu_{21} \over \sqrt{\mu}}    
\right)^{-1}  \left\{    
\left({2\mu_{23} \over     
\sqrt{\mu}}\right)^{\sum_i k_i} -     
\left({2\mu_{13} \over \sqrt{\mu}}\right)^{\sum_i k_i}    
\right\}. } 
 
Putting everything together, we arrive at the three-point 
function (with $\beta_i=1-k_i$): 
\eqn\threeptfn{\eqalign{    
C^{\mu_2\mu_3\mu_1}_{\b_1, \b_2, \b_3} &= \CB_1\,\CB_2\,\CB_3\,\CB_4\cr    
& = {(-1)^{\lfloor\Sigma_i k_i/2\rfloor} \over   
(2\pi\epsilon)^{1+\Sigma_i k_i}}    
{\mu_{23}^{\Sigma_i k_i} - \mu_{13}^{\Sigma_i k_i}\over\mu_{21}}   
{\G_1\left(1+\sum_i k_i\right) \over \G_1(2)} \prod_{j=1}^3  
{\G_1\left(1+\sum_i k_i-2k_j\right) \over \G_1(2k_j)}.  
}}  
In terms of the renormalised operators defined in Eq.\renfield\ and 
\renLiou, the three-point function becomes: 
\eqn\threeptfnren{ 
\t C^{\mu_2\mu_3\mu_1}_{\b_1, \b_2, \b_3} 
 = {(-1)^{\lfloor\Sigma_i k_i/2\rfloor} \over 2\pi\epsilon} 
{\mu_{23}^{\Sigma_i k_i} - \mu_{13}^{\Sigma_i k_i}\over\mu_{21}}   
{\G_1\left(1+\sum_i k_i\right) \over \G_1(2)} \prod_{j=1}^3  
{\G_1\left(1+\sum_i k_i-2k_j\right) \Gamma(2k_j) \over \G_1(2k_j)}. } 
In the special case of tachyons, the momenta of the three operators  
obey $k_1+k_2=k_3$, and the three point function takes the simpler form    
\eqn\renthreetach{ 
\eqalign{    
{\t C}^{\mu_2\mu_3\mu_1}_{\b_1, \b_2, \b_1+\b_2-1} =  
{(-1)^{\lfloor\Sigma_i k_i/2\rfloor} \over \epsilon}\;  
{\mu_{23}^{2k_1+2k_2} - \mu_{13}^{2k_1+2k_2}\over\mu_{21}}.   
}}  
Like the boundary reflection amplitude Eq.\rentwopt, the boundary 
three-point function obtained here also turns out to be independent of 
the bulk cosmological constant, depends only on pairwise differences 
of boundary cosmological constants, and has a simple pole singularity 
in $\epsilon$. We conjecture that these three properties also hold for 
all $n$-point functions of boundary operators in this theory. 
 
As a check, we consider the three-point function with momenta $k_1=k$, 
$k_2=1$ and $k_3=k$ for the three operators. In this case the middle 
operator has $\beta=0$ and hence, if we choose $\sigma_1=\sigma_3$ 
(which implies $\mu_{1B}=\mu_{3B}$), it reduces to the identity.  Now 
the above correlator should reduce to the two-point function. {}From 
Eq.\threeptfn\ we find: 
\eqn\keqone{ 
C^{\mu_2\mu_1\mu_1}_{1-k, 1, 1-k} = {(-1)^{k} \over  
(2\pi\epsilon)^{2k+2}}\, 
{2\pi\,\mu_{21}^{2k}\over \Gamma(2k+1)\Gamma(2k)}.}  
Comparing with Eq.\twopointfinal, we see that this is related to the 
(bare) reflection amplitude by: 
\eqn\threeptrefl{ 
C^{\mu_2\mu_1\mu_1}_{1-k, 1,1- k} = {1\over 2\pi\epsilon}\,  
d(1-k|\mu_1,\mu_2). } 
If we interpret ${1\over 2\pi\epsilon}$ as the $\delta(0)$ factor 
arising from $\delta(\beta_1-\beta_2)$ in Eq.\twoptbasic, we may 
conclude that in the special case being considered, the three-point 
function indeed reduces to the two-point function as expected.


\newsec{Bulk-Boundary Two-Point Function}   
 
The bulk-boundary two-point function on the disc involves a boundary    
operator $V_\beta^{\s\s}$ and a bulk operator ${\cal V}_\alpha$. This 
was computed in Ref.\refs{\hosomichi} (see also  
Refs.\refs{\kospon,\teschner}) and the result is: 
\eqn\defbb{   
\left\langle {\cal V}_\alpha(z,\bar z)V_\beta^{\s\s}(x)\right\rangle   
= {A_{\alpha\beta}^\s\over |z-\bar z|^{2\Delta_\alpha-\Delta_\beta}\,   
|z-x|^{2\Delta_\beta}}, }   
where,    
\eqn\bbexpr{   
\eqalign{   
A_{\alpha\beta}^\s &= \CC_1\,\CC_2\,\CC_3,\cr   
\CC_1 &= 2\pi\left(\pi\mu_0\gamma(b^2)b^{2-2b^2}\right)^{(Q-2\alpha-   
\beta)/2},\cr   
\CC_2 &= {\Gamma_1^3(Q-\beta)\Gamma_1(2\alpha-\beta)\Gamma_1(2Q-2\alpha   
-\beta)\over\Gamma_1(Q)\Gamma_1(Q-2\beta)\Gamma_1(\beta)\Gamma_1(2\alpha)   
\Gamma_1(Q-2\alpha)},\cr   
\CC_3 &= {1\over i}\int_{-i\infty}^{i\infty} dt\;    
e^{2\pi i(2\s-Q)t}\,   
{S_1\left(t+{1\over2}\beta+\alpha-{1\over2}Q\right)   
S_1\left(t+{1\over2}\beta-\alpha+{1\over2}Q\right)\over   
S_1\left(t-{1\over2}\beta-\alpha+{3\over2}Q\right)   
S_1\left(t-{1\over2}\beta+\alpha+{1\over2}Q\right)}. }}   
We want to evaluate this in the $c=1$ string theory where, as usual, we  
need to take the singular limit $b=1$. Let us also recall that  
$\beta=1-k$ and $\alpha=1-{1\over2}k$ --- the bulk and boundary  
windings are related due to the winding number conservation condition  
from the matter sector. We will assume that $k>0$.  The first factor  
$\CC_1$ can be rewritten using the the by-now familiar renormalized  
bulk cosmological constant:  
\eqn\calcone{   
\CC_1= 2\pi \left({\mu\over 4}\right)^{k-{1\over2}}, }   
while the second factor is easily evaluated to be:   
\eqn\calctwo{   
\CC_2 = {(-1)^{1+k}\over 2\pi\Gamma(2k)\left(\Gamma(k)\right)^2}\;   
(2\pi\epsilon)^{2k-1}. }   
Finally, we come to the third factor $\CC_3$ which involves an integral   
similar to the one encountered in the evaluation of the boundary   
three-point function. Specifically, we have to evaluate   
\eqn\calcthr{   
\CC_3 = {1\over i} \displaystyle\int_{-i\infty}^{i\infty} dt\;    
\exp\left(4\pi i(\s-1)t\right)\,   
{S_1\left(t-k+{1\over2}\right)S_1\left(t+{1\over2}\right)\over   
S_1\left(t+k+{3\over2}\right)S_1\left(t+{3\over2}\right)}. }   

For large imaginary values of $t$, the integrand falls off  
exponentially. This makes the integral \calcthr\ convergent. In the  
kinematic region where $k$ is negative, all the poles of the integrand  
arising from the numerator are in the left half-plane while those from  
the denominator are in the right half-plane.  For other values of $k$  
the integral is defined by analytic continuation described in detail  
in Appendix B. Once again, the integral is dominated by its  
singular part, which comes from the collision of the poles from the two  
half-planes. Denoting $t+{1\over2}=n$, the conditions for collision  
are met for integer values of $n$ between $1-k$ and $k$.  Evaluating  
the (singular) residues at these poles we find  
\eqn\evalint{   
\CC_3 = {1\over(2\pi\epsilon)^{2k+1}}\,e^{-2\pi i\s}\;   
\sum_{n=1-k}^k e^{4\pi i\s n}   
=  {1\over(2\pi\epsilon)^{2k+1}}\,{\sin(4\pi\s k)\over\sin{2\pi\s}}.}    

Combining Eqs.\calcone, \calctwo\ and \evalint, the bulk-boundary two   
point function of tachyons is found to be:   
\eqn\bulkboun{   
\eqalign{   
A_{\alpha\beta}^\s &= \left({\sqrt{\mu}\over2}\right)^{2k-1}\,   
{(-1)^{k-1}\over \epsilon^2\,\Gamma(2k)\left(\Gamma(k)\right)^2}\;   
{\sin (4\pi\s k)\over\sin (2\pi\s)}\cr   
&= \left({\sqrt{\mu}\over2}\right)^{2k-1}\,   
{(-1)^{k-1}\over \epsilon^2\,\Gamma(2k)\left(\Gamma(k)\right)^2}\;   
\sum_{\ell=1}^{k}(-1)^{\ell+1}\left({2k-\ell\atop \ell -1}\right)\,   
\left({2\mu_B\over\sqrt{\mu}}\right)^{2k-2\ell+1}. }}   
Like the previous correlators, this too is conveniently expressed in  
terms of renormalised bulk and boundary operators, the latter being  
given by Eqn.\renfield\ and for the former we choose:  
\eqn\renbulk{  
\t Z(k;m,\bar m) = (2\pi\epsilon)^{-k}{\Gamma(k)\over   
\Gamma(1-k)}\, Z(k;m,\bar m). }  
Once again, this redefinition differs from the standard one and is   
chosen so as simplify the form of the renormalized expression.   
Specialising to the tachyons \bulktachvert,\bndrytach\ for simplicity,  
\eqn\bulkbounren{  
\left\langle \t T^{\s\s}(k)\t{\cal T}(k) \right\rangle =   
\left({\sqrt{\mu}\over2}\right)^{2k-1}\,   
{(-1)^{k-1}\over\epsilon^2}{\sin(\pi k)\sin (4\pi\s k)\over  
\pi\sin (2\pi\s)}. }  
Unlike the boundary two- and three-point functions, we see that the 
bulk-boundary correlator does depend explicitly on the bulk 
cosmological constant $\mu$, through $\sigma$. It also lacks the 
translational symmetry in $\mu_B$ that we found in the boundary 
correlators. (This was to be expected, since there is only one boundary 
operator $V_\beta^{\s\s}$ and this is necessarily diagonal in the 
boundary cosmological constant. However, we suspect that with more 
boundary operators too, the bulk-boundary correlators will lack 
translational symmetry in the $\mu_B$.) Finally, we see that this 
correlator has a double pole singularity in $\epsilon$, unlike the 
simple pole found in the boundary correlators. 
 
Specialising further to $k=0$ (the cosmological operators) we find:  
\eqn\bbcosmo{  
\left\langle \t T^{\s\s}(0)\t{\cal T}(0) \right\rangle =  
\left.\left\langle \t T^{\s\s}(k)\t{\cal T}(k)  
\right\rangle\right|_{k\to 0}  
= -\,{2\over\sqrt\mu}\,{4\pi\s\over\sin 2\pi\s}.}  
Interestingly, in this case the correlator is non-singular. 
 
As a consistency check, the bulk-boundary two-point function, if 
correctly normalised, should reduce to the bulk one point function 
when the boundary Liouville momentum vanishes, $\beta\to 0$. This 
corresponds to $k=1$ in our case. The bulk one-point function of 
Liouville theory is given by\refs{\fzz}: 
\eqn\oneptfn{  
\langle {\cal V}_\alpha(z,\zb) \rangle_\s  
= {U_\s(\alpha)\over |z-\zb|^{2\Delta_\alpha}}, }  
where,  
\eqn\oneptnext{  
\eqalign{ 
U_\s(\alpha) &= {2\over  b}  
\left(\pi\mu\gamma(b^2)\right)^{Q-2\alpha\over 2b}\Gamma(2b\alpha-b^2)  
\, \Gamma\left({2\alpha\over b} - {1\over b^2}-1\right)\cr 
&\qquad\times\;  \cos\big(\pi(2\alpha-Q)(2\s-Q)\big). }}  
Due to the momentum conservation condition from the matter sector, 
this should strictly be evaluated only at $k=0$. Nevertheless, let us 
keep $k$ arbitrary at this stage. Putting $b=1$ in \oneptnext\ and 
performing the familiar renormalization of cosmological constants, as 
well as renormalization of the bulk tachyon as in Eq.\renbulk, the 
one-point function is: 
\eqn\oneptresult{  
\t U_\s(k) = -\,\left({\sqrt\mu\over 2}\right)^{k}  
(2\pi\epsilon)^{-k}{2\pi\cos(2\pi\sigma k) \over   
k\sin(\pi k)}. }  
The bulk-one point function above is seen to satisfy the expected  
functional equation  
$\t U_{\s+\half}(k) + \t U_{\s-\half}(k) = 2\cos(\pi k)\,\t U_\s(k)$ 
rather trivially\kospon. Substituting $k=1$ (hence  
$\alpha=1-{k\over2}={1\over2}$) formally,   
\eqn\redundanteqn{  
\t U_\s(k=1) = -\,{\sqrt\mu\over2}\,{\cos 2\pi\s\over\pi\epsilon^2}, }  
On the other hand, Eq.\bulkbounren\ evaluated at $k=1$ gives: 
\eqn\bulkboundaryreln{ 
\left\langle  
\t T^{\s\s}(1,1)\t{\cal T}(1) \right\rangle =   
{\sqrt\mu\over2}\,{2\cos 2\pi\s\over\epsilon}. } 
Recalling from Eq.\renfield\ that ${\t T}(1) = 2\pi \epsilon\, T(1)$ we 
see that 
\eqn\bulkboundaryunrenorm{ 
\left\langle  
T^{\s\s}(1)\t{\cal T}(1) \right\rangle =   
{\sqrt\mu\over2}\,{\cos 2\pi\s\over\pi\epsilon^2}. } 
This agrees with Eq.\redundanteqn\ (upto a sign). 

   
\newsec{Physical Correlators and Open String Field Theory}  
 
According to a recent proposal of 
Sen\refs{\SenOCone\SenOCtwo{--}\SenOCthree}, open-string field theory on 
D-branes in a certain background is dual to a theory of closed strings 
to which the branes in that background couple. In most known examples, 
the complete string field theory is extremely complicated, and lacking 
the necessary analytic tools, is only accessible through approximation 
schemes such as level truncation. Having examples of D-branes on which 
the full open-string field theory can be analysed is clearly 
important.  Non-critical string theories, with their relatively simple 
yet rich physical content and high degree of symmetry, are string 
backgrounds where we may further our understanding of this 
duality\refs{\SenOCtwo,\SenOCthree}. Indeed in Ref.\gaiotto, the 
topological Kontsevich matrix model of topological gravity (equivalent 
to $c=-2$ closed-string theory) is shown to arise from the 
open-string field theory of the branes of $c=-2$ matter coupled to 
Liouville theory. 
 
In this section, we take the first step in this direction for the case 
of the $N$ FZZT branes in the $c=1$ theory compactified at 
$R=1$. Open-string field theory has an infinite number of fields, but 
it also has infinite gauge redundancy. The closed-string sector of the 
$c=1$ theory at the self-dual radius, and indeed of all non-critical 
string theories, possesses a topological symmetry due to which only 
degenerate worldsheets at the boundary of the moduli space of Riemann 
surfaces contribute to correlators. In other words, only physical 
(`on-shell') states in the cohomology of the BRS operator $Q_B$ 
contribute to quantum string amplitudes, at all genus. This is the 
well-known topological localisation. 
 
When D-branes, {\it i.e.\ } open strings, are included, we lack a 
direct proof that this property continues to hold. However, an 
important source of intuition comes from the relation of the bulk 
theory to the topological SL(2)/U(1) coset\sunilvafa\ and the deformed 
conifold\debvafa. There has been progress in understanding 
localisation in the open-string sector of the topological SU(2)/U(1) 
cosets\openloc, closely related to the first description. On the other 
hand, a classic result due to Witten\wittenCS\ tells us that the 
open-string field theory on D3-branes wrapping a 3-cycle of the 
deformed conifold localises to pure Chern-Simons theory (this 
considerably preceded the discovery of D-branes!). With this 
motivation, for the moment we simply assume that the string field 
theory localises onto the physical states (as defined by the BRS 
cohomology) and arrive at an action for these modes, postponing a 
detailed analysis of localisation and the resulting model for future 
work\foot{In \gaiotto, a powerful nilpotent symmetry of the gauge 
fixed quantum action of the $c=-2$ noncritical string theory is 
exploited for localisation. The existence of such a symmetry is 
stronger in that it takes into account the effect of worldsheet 
instantons. We note, however, that the absence of compact two-cycles 
in the deformed conifold geometry will forbid potential instanton 
correction.}. 
 
Let the CFT Hilbert space of the states of the first-quantized string 
between the $i$th and the $j$th brane be ${\cal H}_{ij}$ 
($i,j=1,\cdots,N$). The open-string field $|\Psi_{(ij)}\rangle$ is a 
ghost-number one state in this Hilbert space. The action defining the 
classical string field theory is 
\eqn\Wsftaction{  
S[\Psi] = - {1\over 2g_{s0}}\sum_{ij}\left\langle \Psi_{(ij)},  
Q_B\Psi_{(ji)}\right\rangle - {1\over 3g_{s0}}\sum_{ijk}  
\left\langle \Psi_{(ij)}\Psi_{(jk)}\Psi_{(ki)}\right\rangle, }  
where the quadratic and the cubic terms are given in terms of CFT 
correlators and $Q_B$ is the BRST operator (see \TayZw\ for a review). 
The linearised equation of motion of the theory $Q_B|\Psi\rangle=0$ is 
the statement that the worldsheet configuration is physical in the 
free theory --- the cubic term then describes interactions. 
 
Our task now is to compute the OSFT action \Wsftaction\ on the FZZT
branes localized onto the physical states. This amounts to the
evaluation of correlators in the CFT of $c=1$ matter plus Liouville
plus the $(b,c)$ ghosts. In Liouville theory, there is no sense in
which the string coupling is weak, therefore we cannot really regard
the cubic term as a perturbation. This is reflected in the fact that
there is an infrared divergence in the two-point correlators of the
physical states. We shall use the same regulator that we did earlier
and see that the kinetic and the cubic term, evaluated on the on-shell
states, contribute to the same order.
 
We have seen earlier that the physical states of the background CFT 
are summarized as an $N\times N$ matrix field living on a 2-sphere. 
The expansion of the open-string field in terms of these states is: 
\eqn\strfld{   
\left|\Psi_{(ij)}\right\rangle = \sum_{k,m} T_{ij}(k,m)  
\left|\t Y^{ij}(k,m)\right\rangle, }  
where $|\t Y^{ij}(k,m)\rangle$ is the ghost number one primary state 
in the boundary CFT corresponding to the open string with ends on 
branes $(i,j)$ transforming as spherical harmonics ${\cal Y}_k^m 
(\theta,\phi)$ under rotations of $S^2$. Under the assumptions 
described above, the open-string field theory action reduces to an 
action for these matrices. 
 
The coefficients of the string field in the OSFT action are determined 
by the CFT correlation functions for primary operators. The matter and 
ghost contribution to the two- and three-point correlators are very 
simple. As we saw in section 2, the physical operators behave exactly 
like the spherical harmonics. For the two-point function, the matter 
contribution is the condition $k_1=k_2\equiv k$ and the conservation  
condition $m_1=-m_2\equiv m$ of the $J_3$ component of angular momentum.   
The full two-point function is: 
\eqn\twopttotal{   
D_{ij}(k) \equiv \left\langle {\t Y^{ij}(k,m)}, c_0 {\t Y^{ji}(k,-m)} 
\right\rangle 
= {(-1)^{k} \over \epsilon^2} {(z_j-z_i)^{2k}\over 2k},} 
where, we have introduced the notation $\mu_B\equiv z$. Let us note that 
for the special case of the cosmological operators, both the terms in 
Eq.~\twoptbasic\ contribute.  
 
The full three-point function is determined from the SU(2) addition 
condition from the matter sector and, using Eq.\threeptfnren, its 
expression is: 
\eqn\threepttotal{ 
\eqalign{ 
C_{jki}(k_1,k_2,k_3) &\equiv \left\langle {\t Y^{jk}(k_1,m_1)}  
{\t Y^{ki}(k_2,m_2)} {\t Y^{ij}(k_3,m_3)} \right\rangle  \cr  
& = {(-1)^{\Sigma_i k_i/2} \over (2\pi\epsilon)}    
\left( {z_{jk}^{\Sigma_i k_i} - z_{ik}^{\Sigma_i k_i}\over z_{ji}}  \right) 
{\G_1(1+\Sigma_i k_i) \over \G_1(2)} \prod_{j=1}^3  
{\G_1(1+\Sigma_i k_i-2k_j) \G(2k_j)\over \G_1(2k_j)} \cr 
&\qquad \qquad \times\;\int d(\cos \theta) d\phi\;  
{\cal Y}_{k_1}^{m_1}(\theta,\phi)  {\cal Y}_{k_2}^{m_2}(\theta,\phi)  
{\cal Y}_{k_3}^{m_3}(\theta,\phi),  
}}  
where ${\cal Y}^k_m(\theta,\phi)$ are the spherical harmonics of 
$S^2$.  Let us recall that the three-point function is evaluated with 
the condition $k_3 > k_1, k_2$, a choice made in evaluating the 
Liouville correlators in Sec.~4. We would also like to point out that 
the extra divergence present in the two-point function \twopttotal\ 
compared to the three-point function \threepttotal\ is due to the 
delta function in \twoptbasic, which can be understood as an infra-red 
divergence arising from the volume of the target space. 
 
It is now straightforward to evaluate the action \Wsftaction\ 
localised on the physical states. The kinetic operator $Q_B$ 
simplifies to $c_0L_0$ in the Siegel gauge. This has a zero acting on 
the physical states $|\t Y^{ij}(k,m)\rangle$, (which are dimension zero 
primaries of the underlying CFT): 
\eqn\siegelQB{ 
\eqalign{ 
\left\langle\t Y^{ij}(k',m'), Q_B \t Y^{ji}(k,m)\right\rangle 
&= \left\langle\t Y^{ij}(k',m')\right|c_0 L_0\left|\t Y^{ji}(k,m) 
\right\rangle\cr 
&= \left\langle\t Y^{ij}(k',m')\right|c_0\left(k^2 +\b_2(2-\b_2)- 
1\right)\left|\t Y^{ji}(k,m)\right\rangle\cr 
&= 2k\epsilon\,\left\langle\t Y_{ij}(k',m')\right|c_0\left|\t Y^{ji} 
(k,m)\right\rangle\cr 
&= 2k\epsilon\;D_{ij}(k)\,\delta_{k,k'}\delta_{m+m',0}. 
}} 
This zero absorbs the volume divergence in the two point function.
Using \twopttotal\ in \siegelQB, we get the coefficient of the
non-local kinetic term
\eqn\kinterm{    
\left\langle\t Y^{ij}(k,m), Q_B \t Y^{ji}(k,-m) \right\rangle  
= {(-1)^k \over \epsilon} \left(z_j-z_i \right)^{2k}, } 
which has a simple pole in $\epsilon$. The coefficient of the cubic
term is simply the three-point function \threepttotal\ with the same
singularity. Thus, the two terms in the action \Wsftaction\ have an
identical singular coefficient. Then we can renormalise the string
coupling as
\eqn\gsrenorm{ 
g_s \equiv \epsilon g_{s0}, } 
to get a sensible matrix theory with a finite action. The novelty of 
this matrix model, compared to the existing ones in the literature, is 
that it has the $SU(2)$ symmetry of the theory manifest from the 
beginning. 

Let us note that the (singular) renormalisation \gsrenorm\ of the string
coupling was also necessary in Ref.\gaiotto\ in order to get the Kontsevich
model. It is actually implicit in \gaiotto, where the $1/\epsilon$ 
singularity of the three-point function as well as the delta-functions
in the two-point functions have been suppressed\GRprivate. 
 
The complete action involving all the modes in \strfld\ is a little 
cumbersome. However, if we restrict to the tachyons (thereby giving up 
$SU(2)$ symmetry), we find the action ${\cal S}={\cal S}_2 + {\cal 
S}_3$, where 
\eqn\tachmatrix{ 
\eqalign{ 
{\cal S}_2 &= -{1\over 2 g_s}\sum_{k=0}^\infty(-1)^k\sum_{ij}  
\t T^{ij}(k)\left(z_j-z_i\right)^{2k}\t T^{ji}(-k)\cr 
{\cal S}_3&= -{1\over 3 g_s}\sum_{k_1,k_2}(-1)^{k_1+k_2} 
\sum_{ijl}{z_{jl}^{2k_2+2k_2}-z_{il}^{2k_2+2k_2}\over z_{ji}} 
\t T^{jl}(k_1)\t T^{li}(k_2)\t T^{ij}(-k_1-k_2). }} 
In terms of a matrix $Z=\hbox{\rm diag}\,(iz_1,iz_2,\cdots)$, the
kinetic term may be written as
\eqn\kintermmat{ 
{\cal S}_2 \sim 
\sum_k\hbox{\rm Tr}\;\t T(k)\left[Z,\cdots\left[Z,\t T(k)\right] 
\cdots\right].} 
The fact that our Liouville correlators depend only on the difference 
of boundary cosmological constants shows up as a symmetry of the above 
term under a shift of the matrix $\t T$ by an arbitrary diagonal 
matrix. This symmetry is shared by the cubic term which can also 
be written down similarly. 
 
 
\newsec{Discussion} 
 
We have studied correlators of the boundary Liouville theory in the
limit that the Liouville central charge $c_L$ tends to 25, or
equivalently $c\to 1$. The results are embodied in
Eqs.\twopointfinal{--}\rentwopt, \threeptfnren{--}\renthreetach\ and
\bulkbounren. The principal motivation to present these results is
that they are far more explicit than the boundary correlators known
for the $c<1$ theory (as embodied in Eqs.\twoptfactors,\defBall\ and
\bbexpr). The latter are given in terms of special functions
$S_b(x),\Gamma_b(x)$ and some of the correlators are known only as
contour integrals over products of such functions. These contour
integrals can be explicitly evaluated for $c=1$ only, as far as we
know, at the self-dual radius\foot{Of course, rational multiples of
this radius which correspond to orbifolds of the theory also have a
similar behaviour.}.  The boundary correlators we obtain in this way
are all divergent, but as we have noted, the divergence factors out
from the two- and three-point functions and can be absorbed in a
rescaling of the string coupling leading to a well-defined open-string
field theory action.
 
The fact that the boundary correlators are independent of the bulk
cosmological constant is reminiscent of a similar fact in
Ref.\refs\gaiotto. There, the dependence of the two-point function on
$\mu_B$ is crucial in recovering the Kontsevich model\Kontsev, where
the different $\mu_{B,i}$ turn into the eigenvalues of the Kontsevich
matrix. In similar vein, our matrix model depends only on $\mu_{B,i}$
which are the eigenvalues of a constant matrix $Z$. 

We did not find a proof that the boundary correlators at $c=1$ and
selfdual radius are all independent of the bulk cosmological
constant. However, if we assume this to be true, then we can see that
the $n$-point tree-level boundary correlators must scale with the boundary
cosmological constant $\mu_B$ as:
\eqn\corrscaling{
\langle V(k_1)V(k_2)\cdots V(k_n)\rangle \sim \mu_B^{\sum_{i=1}^n k_i -
n +2} }
where a factor of $(k_i-1)$ comes from each Liouville vertex operator
and an additional 2 comes from the linear dilaton factor in the path
integral. This scaling is satisfied by the two- and three-point
correlators that we computed. It is tempting to also conjecture that
the $n$-point correlators will depend only on the pairwise differences
$\mu_{ij}$ of boundary cosmological constants. 

The natural matrix model that we might have expected to find from our
computations, which is the analogue of the Kontsevich model for $c=1$
at self-dual radius, is the model of Ref.\refs\IM. But this is a
one-matrix model, and here we find a model with infinitely many
matrices. Moreover the model of
\refs\IM\ incorporates amplitudes for (closed-string) tachyon
external states only, based as it is on the amplitudes computed in
Ref.\refs\DMP\ from matrix quantum mechanics, in which the other
discrete states have not yet been constructed. So there is in fact no
candidate matrix model presently available that incorporates the full
$SU(2)$ symmetry of the $c=1$ string at self-dual radius. In contrast,
the approach in the present paper does lead to such a model, presented
in embryonic form in Eqs.{\threepttotal{--}\kinterm} More work is
needed to understand this model and confirm whether open/closed
duality works as expected.
 

 
\noindent{\bf Acknowledgements:}
\smallskip

\noindent D. Ghoshal and S. Murthy would like to thank  
the Tata Institute of Fundamental Research, and S. Mukhi is grateful
to the University of Amsterdam and the Les Houches School on Random 
Matrices, 
for their hospitality while parts of this work were carried
out. It is a pleasure to thank Sergei Alexandrov, Chris Beasley,
Sergey Cherkis, Davide Gaiotto, Simeon Hellerman, Camillo Imbimbo, 
Nathan Seiberg and
especially John McGreevy and Leonardo Rastelli for very useful
discussions.  The research of D. Ghoshal is supported in part by an
Invitation Fellowship of the Japan Society for the Promotion of
Science (JSPS), and that of S. Murthy by the NSF Grant
PHY-0243680. Any opinions, findings, and conclusions or
recommendations expressed in this material are those of the authors
and do not necessarily reflect the views of the National Science
Foundation.
 

   
\appendix{A}{Special Functions at $c=1$}   
 
The correlators of Liouville theory are expressed in terms of some
special functions\refs{\fzz,\teschner}. In the case of $c_L=25$, {\it
i.e.}, $b=1$, they are:
\eqn\spfnasintegral{   
\eqalign{   
\ln\Gamma_1(x) &= \int_0^\infty {dt\over t}\left({e^{-xt}-e^{-t}\over   
(1-e^{-t})^2} - {(1-x)^2\over 2e^t} - {2(1-x)\over t}\right),\cr   
\ln S_1(x) &= \int_0^\infty {dt\over t}\left({\sinh 2t(1-x)\over   
2\sinh^2 t} - {1-x\over t}\right). }}   
Both are meromorphic functions and are related to each other via:   
\eqn\SrelGamma{   
S_1(x) = {1\over S_1(2-x)} = {\Gamma_1(x)\over\Gamma_1(2-x)}, }   
where, we have also made use of the unitarity relation 
$S_1(x)S_1(2-x)=1$. The function $\Gamma_1$ has poles at zero and 
negative integer arguments. Therefore, from Eq.\SrelGamma, $S_1(x)$ 
has poles at these arguments and zeroes at integers larger than 1. 
   
The functions $\Gamma_1(x)$ and $S_1(x)$ satisfy the recursion    
relations:   
\eqn\GammaRR{   
\eqalign{   
\Gamma_1(x+1) &= {\sqrt{2\pi}\over\Gamma(x)}\;\Gamma_1(x),\cr   
S_1(x+1) &= 2\sin(\pi x)\,S_1(x); }}   
where, $\Gamma(x)$ is the usual Euler gamma function.    
The values of these special functions at (half)-integer arguments   
turn out to be of interest. In particular, we would need the ratio   
$\Gamma_1(-n)/\Gamma_1(n)$, which, as a matter of fact, is   
divergent. However, using the recursion relations above, one can    
show that the leading divergence, near an integer $n$ is   
\eqn\ratioGint{   
\eqalign{   
{\Gamma_1(-n)\over\Gamma_1(n)} &\equiv \lim_{\epsilon\to 0+}   
{\Gamma_1(-n-\epsilon)\over\Gamma_1(n+\epsilon)}\cr   
&= {(-1)^{n(n+1)/2}\over(2\pi)^n   
\Gamma(n)\Gamma(n+1)}\;{1\over\epsilon^{n+1}}. }}   
However, for half-integer arguments:   
\eqn\ratioGhalfint{   
{\Gamma_1\left(-{2m+1\over2}\right)\over\Gamma_1\left({2m+1\over   
2}\right)} = {(-1)^{(m+1)(m+2)/2}\sqrt{2}\over    
\pi^{m+{3\over2}}\,(2m-1)!!\,(2m+1)!!},\quad m\in\hbox{\bf Z}, }   
the corresponding ratio is finite.    
   
Likewise, using the relations above, $S_1(1-x)={\Gamma(x)   
\Gamma_1(-x)\over\Gamma(-x)\Gamma_1(x)}$. Therefore, one finds   
that   
\eqn\Satint{   
\eqalign{   
S_1(1-n) &= {(-1)^{n(n-1)/2}\over (2\pi\epsilon)^n},\cr   
S_1\left(1-{2n+1\over2}\right) &= {(-1)^{-n(n+1)/2}   
\over 2^{2n+{1\over2}}\pi^{n+{3\over2}} }, }}   
for an integer $n$. Once again, the first of the above is to   
be defined as a limit.     
 

\appendix{B}{Evaluation of a Contour Integral for the Three-Point 
Function} 
 
Here we will evaluate the contour integral 
\eqn\contourint{ 
\CB_4 = {1 \over i} \int_{-i \infty -0}^{+i \infty -0} ds \;\;    
\prod_{i=1}^4{S_b(U_i+s) \over S_b(V_i+s)} } 
where $U_i,V_i$ are given in Eq.\defuv.  The definition of the contour
integral as an analytic function of the momenta is explained in
Ref.~\teschner. We shall summarize and use that prescription for our
case in which $k_i$ approach positive integers, and $b \to 1$.  We
shall use an off-shell parameter $\epsilon$ here which is the
deformation $b$ away from $b=1$. As mentioned in section 3, an
equivalent deformation is one where the Liouville momenta is shifted
away from integers.

For large imaginary $|s|$, the integrand decays exponentially, so the 
integral is convergent in that region. Near the origin, the contour 
needs to be defined because the integrand has poles which lie on the 
origin. We do this by shifting the contour a little to the left of the 
imaginary axis. 
   
Let us list the arguments of the functions $S_1$ for our case:   
\eqn\defuvbone{\eqalign{    
U_1 =-1+\sigma_1+\sigma_2+k_1, \qquad  & V_1 = 2+\sigma_2-    
\sigma_3+k_1+k_3, \cr    
U_2 = 1-\sigma_1+\sigma_2+k_1, \qquad & V_2 = 2+\sigma_2-\sigma_3+  
k_1-k_3,\cr    
U_3 = 1+\sigma_2-\sigma_3-k_2, \qquad & V_3 = 2\sigma_2, \cr    
U_4 = 1+\sigma_2-\sigma_3+k_2, \qquad & V_4 = 2.\cr }}   
The poles from the numerator and the denominator are at\foot{Here we  
have already plugged in $b=1$, the general formula has simple poles at  
$s+U_i=-nb-mb^{-1}$. At $b=1$, these simple poles  
coalesce to a pole of high order. 
}  
\eqn\poles{    
s+U_i=-n_i\quad\hbox{and }\quad     
s+V_i=2+m_i,\qquad (n_i,m_i=0,1,2,\cdots) }    
respectively.  For Re~$U_i > 0$ and Re~$V_i \le 2$, the poles arising
from the numerator are all in the left half-plane and those from the
denominator are in the right half-plane\foot{The $V_4$ factor has a
pole at the origin, but we have shifted the contour a little to the
left as indicated in \defBall. With this understanding, we shall
continue to call it the imaginary axis.}. The imaginary axis is
therefore a well-defined contour and thanks to the asymptotic
behaviour, the integral has a finite value.
    
For general values of $k_i$, the integral is defined by analytic   
continuation of the above prescription. Specifically, this means that   
as we vary $k_i$ (or equivalently, $U_i, V_i$) smoothly, some of the   
poles from the LHP cross the imaginary axis and enter the RHP, and   
vice-versa. In such a case, one deforms the contour such that the   
poles from the numerator and the denominator are always separated by   
the contour. Alternatively, this could be done by an equivalent   
deformation as follows. Suppose, a pole of the numerator migrates to   
the LHP. The new (deformed) contour now consists of two parts, one is   
the old one and another a small circle around the `migrating'   
pole. The latter will pick up the residue of the integrand around that   
pole. However, this also gives a finite contribution and will not be   
of our final interest.   
   
The integral diverges if two poles, one originating in numerator and
another in denominator, approach towards each other to coincide. In
this case, the contour is `pinched between' the two
poles. Alternatively, the migrating pole hits another pole. This
divergence dominates over the finite piece and it is this which is of
interest to us. In order to extract the leading divergence in such
cases, let us deform $b$ away from the value $b=1$ by an amount
$\epsilon$ and make the circle around a migrating pole very small. As
it hits a would-be singularity at $b=1$, we determine the divergent
residue as a power of $\epsilon$.
    
The condition for collision between the poles \poles\ is   
$s = -U_i-n_i=2-V_j+m_j$, ($n_i,m_j=0,1,2,\cdots$), {\it i.e.},    
\eqn\collision{   
V_j-U_i=2+m,\; m=0,1,2,\cdots.   
}    
For generic $\s_i$, this can only happen when $V_1$ collides with    
$U_3$ or $U_4$. Moreover, $V_1-U_3=1+k_1+k_2+k_3=V_1-U_4+2k_2$, so the     
divergence from the collision of $V_1$ and $U_3$ dominates and   
it is sufficient to consider only that. Let    
\eqn\defn{    
s+(\s_2-\s_3)=n \in \IZ.    
}    
Then, a collision between the poles \poles\ happens when    
\eqn\speccoll{    
1+n-k_2=-n_3, \qquad n + k_1 +k_3 = m_1, \qquad (n_3,m_1 = 0,1,2,\cdots).    
}    
This happens when $-k_1-k_3 \le n \le k_2 -1$. This set is    
non-empty for $(k_1,k_2)\neq (0,0)$.    
 
The divergence of the integrand for a particular value of $n$, as  
defined in Eq.\defn\ above, contributes an amount to the integral 
$\CB_4$ that we denote $\CB_4^{(n)}$. Hence, 
\eqn\bfoursum{ 
\CB_4 = \sum_n \CB_4^{(n)} } 
The range of values of $n$ over which the sum is to be performed will be 
determined below. 
   
The net order of divergence of the integrand at a given value of $n$    
comes from counting the poles/zeroes in $U_3,U_4,V_1$ and $V_2$ (keeping    
$\s_i$ are generic), and (using Eq.\defn\ and the formula     
for the divergence of the $S_1$-function given in Appendix A)  
is equal to:  
\eqn\orderpole{    
- (n-k_2)-(n+k_2) + (1+n+k_1+k_3) + (1+n+k_1-k_3) = 2 + 2k_1. }   
One of these poles is the migrant one with a circular contour around
it, so the divergent part of the residue\foot{Let us see how the same
result is obtained with the equivalent regulator in which $b=1$ but
$k$ is shifted away from an integer. The contour integral is about a 
pole of higher order, say $M\equiv n+k_1+k_3$, if the migrant pole is 
from $V_1$. The residue is then the $(M-1)$th derivative of the other 
factor which has a pole of order $2k_1+2-M$. The dominant singularity
comes from differentiating this singular part, leading to the same final 
answer.} is $1/(2\pi\epsilon)^{2k_1+1}$.
 
The finite piece of the residue is due to the other four $S_1$-functions.    
Once again, using \defn, we can write the contribution $\CB_4^{(n)}$ as: 
\eqn\Bfourmore{    
\CB_4^{(n)}= {(-1)^{k_1} \over (2\pi \epsilon)^{2k_1+1}} {S_1(-1+n+k_1+     
\sigma_1+\sigma_3)\, S_1(1+n+k_1-\sigma_1+\sigma_3) \over      
S_1(n+\sigma_2+\sigma_3)\, S_1(2+n-\sigma_2+\sigma_3)}. }    
Combining $\CB_3$ and $\CB_4^{(n)}$ and using some inversions of $S_1$ along  
the way,  
\eqn\Atwothree{\eqalign{    
\CB_3 \,\CB_4^{(n)} = & {(-1)^{k_1} \over (2\pi \epsilon)^{2k_1+1}} \;    
{S_1(-1+n+k_1+ \sigma_1+\sigma_3) \over S_1(-1-k_3+\sigma_1+\sigma_3)}     
{S_1(-1+k_2+\sigma_2+\sigma_3) \over S_1(n+\sigma_2+\sigma_3)} \cr    
& \qquad \qquad \times{S_1(1+k_3+\sigma_1-\sigma_3)     
\over S_1(1-n-k_1+\sigma_1-\sigma_3)}    
{S_1(-n+\sigma_2-\sigma_3) \over S_1(1-k_2+\sigma_2-\sigma_3)} \cr    
& = {(-1)^{k_1} \over (2\pi \epsilon)^{2k_1+1}}\;   
(2\sin{\pi(\s_1+\s_3)})^{k_1+k_3+n}     
(2\sin{\pi(\s_2+\s_3)})^{k_2-n-1} \cr    
& \qquad \times (2\sin{\pi(\s_1-\s_3)})^{k_1+k_3+n}     
(2\sin{\pi(\s_2-\s_3)})^{k_2-n-1} \cr     
&={(-1)^{k_1} \over (2\pi \epsilon)^{2k_1+1}}\;   
\left(2\,{\mu_{1B}-\mu_{3B} \over     
\sqrt{\mu}}\right)^{k_1+k_3+n} \left(2\,{\mu_{2B}-\mu_{3B}     
\over \sqrt{\mu}}\right)^{k_2-n-1}.    
}}   
 
Finally we have to sum over all these residues, since the contour is a
disjoint sum of all these circles at various values of $s$ labelled by
an integer $n$, which ranges from $-k_1-k_3$ to $k_2-1$.  This is a
geometric series. Evaluating the sum, we get:
\eqn\Atwothreemore{    
\eqalign{ 
\CB_3 \,\CB_4 &= \CB_3\sum_{n=-k_1-k_3}^{k_2-1}\CB_4^{(n)} \cr 
&= {(-1)^{k_1} \over (2 \pi \epsilon)^{2k_1+1}}     
\left({2\mu_{21} \over \sqrt{\mu}}    
\right)^{-1}  \left\{    
\left({2\mu_{23} \over     
\sqrt{\mu}}\right)^{\sum_i k_i} -     
\left({2\mu_{13} \over \sqrt{\mu}}\right)^{\sum_i k_i}    
\right\}, }} 
where we have defined:    
\eqn\muijdef{    
\mu_{ij} \equiv \mu_{iB}-\mu_{jB}. }    
%

\listrefs    
    
\end